\documentclass[fleqn,usenatbib]{mnras}
\usepackage{newtxtext,newtxmath}
\usepackage[T1]{fontenc}
\usepackage{url}
\usepackage{graphicx}
\usepackage{amsmath}
\usepackage{float}
\let\cite\citep
\selectfont

\title[Local bandpower diagnostics for E-mode HPA]{
\boldmath Local bandpower diagnostics for the hemispherical power asymmetry in CMB E-mode polarization}
\author[Robert A. Lynch]{
Robert A. Lynch,$^{1}$\thanks{E-mail: r.a.lynch@gmail.com}
\\
$^{1}$Independent Researcher, Miami, FL 33132, USA
}
\date{Accepted 2026 July 7. Received 2026 July 6; in original form 2026 March 26}
\pubyear{2026}

\begin{document}
\label{firstpage}
\pagerange{\pageref{firstpage}--\pageref{lastpage}}

\maketitle
\begin{abstract}
We present local bandpower diagnostics that test whether a candidate dipolar modulation in E-mode polarization data originates in the cosmic microwave background (CMB) itself rather than in noise or systematics. The target is the hemispherical power asymmetry (HPA), one of the most persistent large-scale anomalies in the CMB temperature field; polarization offers an independent probe of its origin, but at Planck sensitivity the polarized sky is noise-dominated. Local power is estimated in overlapping patches on the masked sky, with E-to-B leakage corrected through standard pseudo-$C_\ell$ deconvolution, and dipole directions are fitted independently to the two Planck half-mission maps, which share the same sky but carry independent noise. Tight directional agreement indicates a CMB-level signal, although it cannot distinguish a real modulation from a chance CMB fluctuation. Applied to Planck PR3 Commander data at $\ell = 15$--$40$, the half-mission directions are separated by $8.7^\circ$, a closer agreement than in any of the 300 isotropic simulations ($p < 0.33\%$), and lie near the temperature HPA axis. The coherence appears only in EE, vanishes without leakage correction, and is corroborated by a cross-spectrum check. LiteBIRD, with substantially higher polarization sensitivity, is a natural next application.
\end{abstract}
\flushbottom

\section{Introduction}
\label{sec:introduction}

The assumption of statistical isotropy --- that the statistical properties of the cosmic microwave background (CMB) are invariant under rotations --- is a foundational prediction of the standard $\Lambda$CDM cosmological model, supported by the simplest inflationary scenarios \cite{Planck2018X}. Observations by WMAP \cite{Bennett2013} and Planck \cite{Planck2018I} have confirmed this prediction to remarkable precision across most angular scales. A number of unexpected features on large angular scales have nevertheless been identified in the temperature fluctuation field, collectively termed ``CMB anomalies'' \cite{Schwarz2016,Planck2018VII}. Among these are the hemispherical power asymmetry \cite{Eriksen2004,Hansen2004}, the alignment of the quadrupole and octopole \cite{deOliveiraCosta2004}, and the Cold Spot \cite{Vielva2004,Cruz2005}.

The hemispherical power asymmetry (HPA) is one of the most persistent and well-characterised of these anomalies. First identified in the first-year WMAP data \cite{Eriksen2004,Hansen2004}, the HPA manifests as a dipolar modulation of the local CMB power spectrum: one hemisphere of the sky exhibits systematically more power than the other. The anomaly has been confirmed across WMAP data releases \cite{Hoftuft2009} and in Planck data \cite{Akrami2014,Planck2018XVI,Planck2018VII}, with a statistical significance exceeding $3\sigma$ in temperature. The preferred direction is consistently found near $(l, b) \approx (221^\circ, -22^\circ) \pm 31^\circ$ (\citealt{Planck2018VII}, QML analysis at $\ell = 2$--$64$) in galactic coordinates. The amplitude of the dipolar modulation is scale-dependent, decreasing with multipole and apparently vanishing above $\ell \sim 600$ \cite{Hoftuft2009,Hanson2009,Planck2018VII}. Separately, \citet{GimenoAmo2025} found that temperature bandpower dipole directions at multipoles up to $\ell \sim 2000$ align with the large-scale HPA axis, a distinct effect from the dipolar modulation measured at $\ell \lesssim 64$.

Temperature measurements have reached the cosmic variance limit at the scales relevant to HPA. The E-mode polarization of the CMB provides a largely independent probe of the same primordial fluctuations. If the HPA originates from a modulation of the primordial power spectrum --- as proposed in several inflationary scenarios~\cite{Erickcek2008} --- a corresponding asymmetry should appear in the E-mode polarization, although the temperature and polarization transfer functions differ and may map the signal to different angular scales~\cite{Ghosh2020}. For a scalar modulation, the expected signal is concentrated in EE; BB is zero at leading order~\cite{Namjoo2015} and affected only at second order with amplitudes equivalent to $r \simeq 0.005$~\cite{Khodagholizadeh2023}; EB vanishes by parity conservation~\cite{Grain2012}. Future experiments with improved polarization sensitivity, notably LiteBIRD~\cite{LiteBIRD2022}, are expected to provide definitive measurements.

Several analyses have searched for the HPA in E-mode polarization. \citet{GimenoAmo2023} applied the local variance estimator (LVE) to Planck PR3 and PR4 E-mode maps after inpainting the masked regions with constrained Gaussian realisations, reporting p-values of $0.22\%$ for SEVEM on PR3 data with a direction of $(l, b) = (232^\circ, -9^\circ)$, and $2.8\%$ on PR4. The significance varies across inpainting realisations, reflecting the stochastic nature of the approach. \citet{Aluri2017} applied the LVE to Planck 2015 Commander E-mode maps and found a dipole modulation amplitude of $2.6$--$3.9\%$, though with a direction pointing toward the kinematic dipole rather than the temperature HPA direction. In a subsequent analysis, \citet{GimenoAmo2025} applied the pseudo-$C_\ell$ estimator in 12 independent sky regions to test whether bandpower dipole directions across multipole bins align with a common axis; no significant alignment was found in E-mode polarization. \citet{Ghosh2020} analysed the squared polarization amplitude $|P|^2 = Q^2 + U^2$ in pixel space, but found that correlated noise and foreground residuals dominated the Planck polarization signal, precluding a detection. They did, however, identify the importance of excluding multipoles below $\ell \sim 10$ for reliable direction recovery.

A fundamental challenge for polarization analyses on incomplete skies is E--B leakage: on a masked sky, the decomposition of $Q$ and $U$ into E-mode and B-mode components is ambiguous, and power leaks between the two~\cite{Lewis2002}. The pseudo-$C_\ell$ framework~\cite{Hivon2002,Alonso2019} addresses this by computing power spectra directly from the masked data and deconvolving the mask-induced mixing through an analytically computed coupling matrix, recovering the true E-mode power without constructing an intermediate E-mode map.

In this work, we apply the same standard pseudo-$C_\ell$ estimator as \citet{GimenoAmo2025}, as implemented in \textsc{NaMaster}~\cite{Alonso2019}, to overlapping patches on the masked sky for dense spatial sampling. The dense patch configuration enables two diagnostics that complement existing amplitude-based analyses. The first is a half-mission directional consistency check: we fit dipoles independently to the HM1 and HM2 bandpower maps and measure the angular separation between the recovered directions. This test cannot establish anisotropy on its own --- a chance dipolar realisation in the shared CMB component produces the same signature as a real modulation --- but it informs whether a candidate signal originates in the CMB rather than in noise or instrumental systematics. The second is an injection-based comparison of the four Planck component-separation pipelines (Commander, SEVEM, SMICA, NILC), which identifies Commander as the most responsive to injected modulations and selects it as the fiducial pipeline. Together with the spectral decomposition --- which yields independent EE, BB, and EB directions from the same deconvolution --- and a cross-spectrum (HM1$\times$HM2) check in which the independent noise terms cancel, these diagnostics address whether a candidate polarization HPA signal originates in the CMB rather than in noise, systematics, or leakage artefacts. These directional diagnostics rely on fitting a dipole independently within each data split, which benefits from dense spatial sampling.

The paper is organized as follows. Section~\ref{sec:data} describes the Planck data and simulations. Section~\ref{sec:method} describes local bandpower estimation, the normalization procedure, the directional consistency check, and the deconvolution ablation test. Section~\ref{sec:validation} validates the pipeline on temperature data and selects the fiducial pipeline through four-method injection comparison. Section~\ref{sec:results} reports the E-mode polarization results, the cross-spectrum check, and the signal injection characterization. Section~\ref{sec:robustness} presents robustness tests. Section~\ref{sec:discussion} discusses the results in context and characterises the test's limitations. Section~\ref{sec:conclusions} summarizes our conclusions.

\section{Data}
\label{sec:data}

We analyse the Planck 2018 (PR3) half-mission data splits produced by four component-separation pipelines: Commander, SEVEM, SMICA, and NILC~\cite{Planck2018IV}. Commander is adopted as the fiducial pipeline, selected through the injection-based comparison described in section~\ref{sec:injection_comparison}; the remaining three methods serve as cross-checks. The half-mission maps (HM1 and HM2) share the same sky signal but have independent noise realisations. For polarization we use the Stokes $Q$ and $U$ maps; for temperature the intensity map, with monopole and dipole removed. We also construct a cross-map as the average of the two half-mission maps: $Q_{\rm cross} = (Q_{\rm HM1} + Q_{\rm HM2})/2$ and likewise for $U$ and $T$. All maps are analysed at HEALPix $N_{\rm side} = 64$~\cite{Gorski2005}, smoothed with a Gaussian beam of FWHM $= 160'$, matching the resolution used in previous large-scale polarization HPA analyses~\cite{GimenoAmo2023}. The beam window function $B_\ell$ and pixel window function $w_\ell$ are applied during map preprocessing; the pseudo-$C_\ell$ estimation and coupling matrix deconvolution operate on the smoothed, pixelized maps. We apply the Planck 2018 confidence masks for temperature and polarization~\cite{Planck2018IV}; the polarization mask retains approximately $78\%$ of the sky.

For calibration and significance assessment we use 300 simulations from the Planck Full Focal Plane 10 (FFP10) suite~\cite{Planck2016XII}, processed through each of the four component-separation pipelines (300 simulations per method). Each simulation consists of a CMB realisation drawn from the Planck best-fit $\Lambda$CDM power spectrum, convolved with the beam and pixel window functions, plus independent half-mission noise realisations matching the noise properties and scanning strategy of the real data:
\begin{equation}
\label{eq:sim_construction}
d_i^{\rm HM1} = B_\ell\, w_\ell\, s_i^{\rm CMB} + n_i^{\rm HM1}, \quad d_i^{\rm HM2} = B_\ell\, w_\ell\, s_i^{\rm CMB} + n_i^{\rm HM2}
\end{equation}
where $s_i^{\rm CMB}$ is the $i$-th CMB realisation (common to both half-missions) and $n_i^{\rm HM1}$, $n_i^{\rm HM2}$ are independent noise realisations. Under the null hypothesis of statistical isotropy, any directional consistency between simulated half-missions arises solely from the shared CMB cosmic variance and not from an anisotropic signal.

\section{Method}
\label{sec:method}

\subsection{Local bandpower estimation}
\label{sec:lbe}

A standard approach to measuring hemispherical power asymmetry divides the sky into patches and assigns a scalar measure of CMB power to each. In the local variance estimator (LVE) used by \citet{Akrami2014} and \citet{GimenoAmo2023}, this scalar is the pixel variance of an E-mode map within the patch. By Parseval's theorem, the pixel variance is equivalent to a weighted sum of the angular power spectrum over all multipoles present in the map:
\begin{equation}
\sigma_i^2 = \sum_\ell \frac{2\ell+1}{4\pi} C_\ell^{\mathrm{patch}}
\label{eq:lve_parseval}
\end{equation}
Equation~(\ref{eq:lve_parseval}) shows the idealised relation; in practice the input maps are beam- and pixel-window-corrected during preprocessing, and auto-spectra include a noise contribution. This is a valid measure of total power, but it has two limitations when applied to E-mode polarization on a masked sky.

First, the sum is broadband, weighted by the number of modes per multipole. For a map at resolution $N_{\mathrm{side}} = 64$, it includes multipoles up to $\ell \sim 191$, well above the range where the dipolar modulation has been detected in temperature ($\ell \lesssim 64$)~\cite{Hanson2009,Planck2018VII}. A broadband sum therefore dilutes the directional signal with isotropic power from scales where no modulation is present.

Second, the LVE requires a global E-mode map as input, from which patches are then extracted. On a masked sky, the decomposition of $Q$ and $U$ into E and B modes is ambiguous: what would be pure E-mode power on the full sky leaks into the B-mode estimate and vice versa. This leakage is determined by the global mask geometry and is present in the E-mode map before any local measurement takes place. The pixel variance within each patch therefore includes leaked B-mode power that the local measurement cannot separate from the true E-mode signal. \citet{GimenoAmo2023} address this through inpainting --- filling masked pixels with constrained Gaussian realisations before decomposing into E and B --- which yields effective leakage correction but depends on the assumed covariance and on the particular realisation used.

The pseudo-$C_\ell$ framework provides a leakage correction that does not require inpainting. Rather than constructing a global E-mode map, the analysis is performed directly on the Stokes $Q$ and $U$ maps within each patch. For each patch, the pseudo-power spectrum $\tilde{C}_\ell$ of the masked $Q/U$ field is computed as a spin-2 quantity. The pseudo-spectrum is related to the true underlying spectra $C_\ell^{EE}$, $C_\ell^{BB}$, and $C_\ell^{EB}$ through a coupling matrix $M_{\ell\ell'}$ computed analytically from that patch's mask geometry. The coupling matrix encodes both the power suppression from incomplete sky coverage and the mixing between E and B modes induced by the mask. Inverting it yields the deconvolved spectra:
\begin{equation}
C_\ell^{EE,\mathrm{patch}} = \left[M^{-1} \tilde{C}\right]_\ell^{EE}
\label{eq:deconvolution}
\end{equation}
We implement this using \textsc{NaMaster}~\cite{Alonso2019}, which computes both the pseudo-spectrum and the coupling matrix for arbitrary spin-2 fields and masks. A single matrix inversion yields the full set of deconvolved EE, BB, and EB power spectra within each patch. \citet{GimenoAmo2025} applied the same pseudo-$C_\ell$ estimator in 12 independent HEALPix sky regions; our application uses 600 overlapping patches for dense spatial sampling.

We compress the EE spectrum to a single scalar by averaging over a chosen multipole range with uniform weight:
\begin{equation}
\hat{C}_i^{EE} = \frac{1}{\ell_{\max} - \ell_{\min} + 1} \sum_{\ell=\ell_{\min}}^{\ell_{\max}} \left[M^{-1} \tilde{C}\right]_\ell^{EE}
\label{eq:bandpower}
\end{equation}
The fiducial range $\ell = 15$--$40$ is motivated by two considerations: the need to exclude the lowest multipoles, which produce cosmic-variance bias in pixel-space estimators --- \citet{Ghosh2020} find that $\ell_{\min} \geq 10$ is sufficient to remove this bias; we adopt the conservative choice $\ell_{\min} = 15$ and verify robustness against $\ell_{\min} = 10$ in section~\ref{sec:robustness} --- and consistency with the angular scales at which the temperature HPA is most significant \citep{Hanson2009}. The temperature and polarization transfer functions differ, so the $\ell$-space mapping of the same primordial signal could differ for E-mode polarization~\cite{Ghosh2020}. Optimization of the $\ell$-range for E-mode sensitivity is deferred to future work. The bandpower averaging also regularizes the coupling matrix inversion: the deconvolution operates on a coarsely binned coupling matrix rather than inverting at individual multipoles, which would be poorly conditioned for the small sky fractions typical of local patches.

We refer to this configuration --- standard pseudo-$C_\ell$ deconvolution applied within local patches and averaged over a chosen multipole range --- as local bandpower estimation (LBE), to distinguish it from the pixel-variance-based LVE in the ablation test of section~\ref{sec:ablation}. The LVE extracts patches from a global E-mode map in which E--B separation has already been performed; the deconvolution performs the E--B separation independently within each patch using that patch's own mask geometry. Deconvolution corrects E--B leakage and mask-induced power suppression; band restriction isolates the angular scales of interest and stabilizes the matrix inversion. Both ingredients are essential: deconvolution without band restriction amplifies noise from poorly conditioned individual multipoles, while band restriction without deconvolution leaves the E--B leakage uncorrected.

\subsection{Patch layout and selection}

Circular patches of radius $R = 15^\circ$ are centred on the pixel centres of a HEALPix grid at $N_{\rm side} = 16$, giving 3072 candidate patch locations. Patches in which at least 60\% of pixels pass the Planck PR3 confidence mask are retained, yielding 2281 candidates for temperature and 2288 for polarization. From these, $N_p = 600$ are selected by greedy distance maximization to ensure approximately uniform sky coverage. The patches overlap, which produces correlations between neighbouring estimates. These correlations do not bias the dipole fit but affect its variance; we account for this through calibration against simulations rather than analytic error propagation.

The patch radius $R = 15^\circ$ is chosen so that modes at $\ell_{\max} = 40$ (corresponding to angular scales of $\sim 4.5^\circ$) fit comfortably within the patch diameter. Alternative radii are tested in section~\ref{sec:robustness}.

The patch mask $w_i(\hat{n})$ for patch $i$ is the product of the galactic mask and a circular boundary:
\begin{equation}
\label{eq:patch_mask}
w_i(\hat{n}) = M_{\rm gal}(\hat{n}) \cdot \Theta(\hat{n} \cdot \hat{n}_i - \cos\alpha_R)
\end{equation}
where $M_{\rm gal}$ is the galactic mask, $\Theta$ is the Heaviside step function, $\alpha_R$ is the angular radius of the patch, and $\hat{n}_i$ is the patch centre. We adopt the unapodised configuration as fiducial, noting that apodisation at the scales tested neither improves nor degrades the result.

\subsection{Normalization}
\label{sec:normalization}

The raw bandpower $\hat{C}_i^{EE}$ in each patch depends on the local noise level, the effective beam, and the patch mask geometry. To isolate the cosmological signal, we subtract the mean bandpower obtained from FFP10 simulations processed through the identical pipeline:
\begin{equation}
\label{eq:normalization}
\hat{C}_i^{\rm norm} = \hat{C}_i^{\rm data} - \langle \hat{C}_i^{\rm sims} \rangle
\end{equation}
where the average is over 300 FFP10 simulations. The mean is computed separately for each half-mission: HM1 data is normalized by the mean of HM1 simulations, and likewise for HM2. This normalization removes the spatially varying baseline due to mask geometry and noise, while preserving any dipolar modulation of the cosmological signal. A patch with excess E-mode power relative to the isotropic expectation will have $\hat{C}_i^{\rm norm} > 0$.

\subsection{Dipole fitting}

A dipole is fit to the normalized bandpower map. The $N_p$ patch centres $\hat{n}_i$ and their normalized bandpowers $\hat{C}_i^{\rm norm}$ are fit to:
\begin{equation}
\label{eq:dipole_model}
\hat{C}_i^{\rm norm} = c_0 + \mathbf{c} \cdot \hat{n}_i
\end{equation}
where $c_0$ is the monopole and $\mathbf{c} = (c_x, c_y, c_z)$ is the dipole vector. The dipole direction in galactic coordinates is $l = \arctan(c_y / c_x)$ and $b = \arcsin(c_z / |\mathbf{c}|)$, and the dipole amplitude is $|\mathbf{c}|$. The dipole is fit with equal weights across all selected patches, with significance calibrated against simulations processed through the identical pipeline.

\subsection{Ablation design}
\label{sec:ablation}

The LVE computes the variance of a reconstructed E-mode map within each patch, without spectral deconvolution or coupling matrix correction. The LBE estimates pseudo-$C_\ell$ spectra from the $Q/U$ maps within each patch and inverts the coupling matrix to obtain deconvolved bandpowers. The two approaches share the same patch layout, normalization, and dipole fitting procedure; they differ in how the per-patch power is estimated and in their treatment of E--B leakage. Comparing them demonstrates that deconvolution changes the recovered signal. The LVE as used by \citet{GimenoAmo2023} incorporates inpainting to mitigate E--B leakage, and its demonstrated sensitivity to the E-mode HPA confirms that some form of leakage correction --- whether through inpainting or through coupling matrix deconvolution --- is the essential ingredient.

\subsection{Half-mission directional consistency check}
\label{sec:consistency_test}

At $\ell = 15$--$40$, the small number of independent modes means that an isotropic CMB realisation produces substantial patch-to-patch power fluctuations, and a dipole fit to these fluctuations yields a nonzero amplitude even under isotropy. For individual half-mission maps, the amplitude of the fitted dipole under isotropy is comparable to the expected HPA modulation at $A = 0.07$, making the half-mission dipole amplitude alone a poor discriminator. The \emph{direction} of the dipole carries additional information. We apply the LBE pipeline independently to HM1 and HM2, obtaining dipole directions $\hat{d}_1$ and $\hat{d}_2$, and quantify their agreement through the angular distance:
\begin{equation}
\label{eq:angular_distance}
\theta = \arccos(\hat{d}_1 \cdot \hat{d}_2)
\end{equation}

The test operates as follows. Each half-mission bandpower map is the sum of a shared CMB realisation and an independent noise realisation. In temperature, where the per-patch signal-to-noise ratio is high, both half-missions recover directions driven by the same underlying CMB realisation; the resulting angular separations are small even under isotropy, leaving little room for a modulation to further tighten the agreement. In E-mode polarization at Planck sensitivity, instrumental noise dominates the per-patch bandpower, scattering the two half-mission dipole directions approximately independently across the sky. If a coherent signal is present in the CMB component --- whether from a real primordial modulation or from a chance dipolar fluctuation in the shared CMB realisation --- it adds a fixed directional anchor to both bandpower maps, pulling both directions toward the same sky location and reducing the angular separation below the isotropic baseline.

This last point requires emphasis. The test cannot establish anisotropy on its own. A chance dipolar realisation in the shared CMB component --- a statistical fluke under isotropy --- produces the same tight half-mission agreement as a real primordial modulation, because both half-missions share the same CMB sky. The test therefore informs whether a candidate signal originates in the CMB rather than in noise or instrumental systematics, but it cannot distinguish a real primordial modulation from a chance CMB fluke. Its role is diagnostic: if the half-mission directions disagree, the candidate signal is likely noise-driven; if they agree, the signal is consistent with a CMB-level origin.

We calibrate the expected distribution of angular separations empirically using 300 FFP10 simulations rather than analytically, since the overlapping patch geometry and non-uniform noise make analytic treatment impractical. The p-value is the fraction of simulations with angular distance smaller than or equal to that observed in the data:
\begin{equation}
\label{eq:pvalue}
p = \frac{1}{N_{\rm sim}} \sum_{j=1}^{N_{\rm sim}} \Theta(\theta_{\rm data} - \theta_j)
\end{equation}

\section{Pipeline selection and validation}
\label{sec:validation}

\subsection{Temperature validation}
\label{sec:tt_results}

The temperature HPA is well established, with significance exceeding $3\sigma$ and a dipole direction of $(l, b) \approx (221^\circ, -22^\circ)$~\cite{Planck2018VII}. We validate the pipeline by applying it to Planck PR3 Commander temperature maps, using identical settings to the polarization analysis --- the same patch selection, the same multipole range ($\ell = 15$--$40$), and the same normalization procedure --- differing only in the use of spin-0 fields (intensity) rather than spin-2 fields ($Q$, $U$).

The LBE recovers a dipole direction of $(l, b) = (223.2^\circ, -36.6^\circ)$ from HM1 and $(223.0^\circ, -36.6^\circ)$ from HM2, an angular separation of $0.1^\circ$ (table~\ref{tab:tt_results}). As discussed in section~\ref{sec:consistency_test}, the high per-patch signal-to-noise ratio in temperature means that both half-missions are dominated by the same CMB realisation; the resulting null distribution is concentrated at small angular separations (simulation mean $0.5^\circ$), leaving little room for a modulation to further tighten the agreement. The recovered direction lies within $14.7^\circ$ of the temperature HPA direction, within the $\pm 31^\circ$ directional uncertainty reported by the Planck Collaboration for the $\ell = 2$--$64$ measurement~\cite{Planck2018VII}.

The LVE applied to the Commander intensity map yields half-mission directions of $(223.1^\circ, -39.5^\circ)$ and $(222.9^\circ, -39.6^\circ)$ with an angular distance of $0.2^\circ$ and $p = 12.3\%$. In temperature, the LVE outperforms the LBE on the directional consistency test, as the pixel-space variance captures the full broadband signal while the LBE restricts to $\ell = 15$--$40$. The distinction between the two approaches emerges in polarization, where E--B leakage correction is required.

\begin{table}
\centering
\caption{Temperature (TT) validation: Commander LBE results at $\ell = 15$--$40$.}\label{tab:tt_results}
\footnotesize
\begin{tabular}{lcccc}
\hline
 & Direction $(l, b)$ & Amplitude & Ang.\ dist. & p-value \\
\hline
HM1 & $(223.2^\circ, -36.6^\circ)$ & $3.26$ & --- & --- \\
HM2 & $(223.0^\circ, -36.6^\circ)$ & $3.25$ & --- & --- \\
HM1--HM2 & --- & --- & $0.1^\circ$ & $6.7\%$ \\
\hline
\end{tabular}
\end{table}

\subsection{Four-method injection comparison}
\label{sec:injection_comparison}

The directional consistency check depends on the component-separation pipeline through the pipeline's treatment of large-scale polarized foregrounds and its noise properties. To select a fiducial pipeline on empirical grounds, we compare the four Planck component-separation methods through signal injection: for each method, a known dipolar modulation $Q(\hat{n}) \rightarrow Q(\hat{n})[1 + A\,\hat{d} \cdot \hat{n}]$ (and equivalently for $U$) is applied to 300 FFP10 simulations at amplitudes $A = 0$, $0.03$, $0.05$, $0.07$, $0.09$, $0.12$, $0.15$, $0.20$, $0.50$, $1.0$, and $2.0$, and the modulated maps are processed through the identical analysis pipeline. The injection direction $(l, b) = (45^\circ, 60^\circ)$ is deliberately offset from the reported HPA direction. Full injection results for the fiducial pipeline are reported in section~\ref{sec:injection}.

The quantity of interest is how each method responds to the injected modulation. The absolute mean HM1$\leftrightarrow$HM2 angular distance at a given amplitude reflects both the method's sensitivity to the modulation and its isotropic baseline --- the mean angular distance at $A = 0$, which differs across methods because of their different noise properties. A method with a large isotropic baseline may show a large absolute reduction in angular distance under injection simply because it started from a higher value, not because it is more sensitive to the underlying signal. The proportional metric identifies the method whose test statistic shifts most in response to a signal, independent of each method's noise baseline under isotropy; this is the operationally relevant quantity for fiducial selection.

\begin{figure}
\centering
\includegraphics[width=\columnwidth]{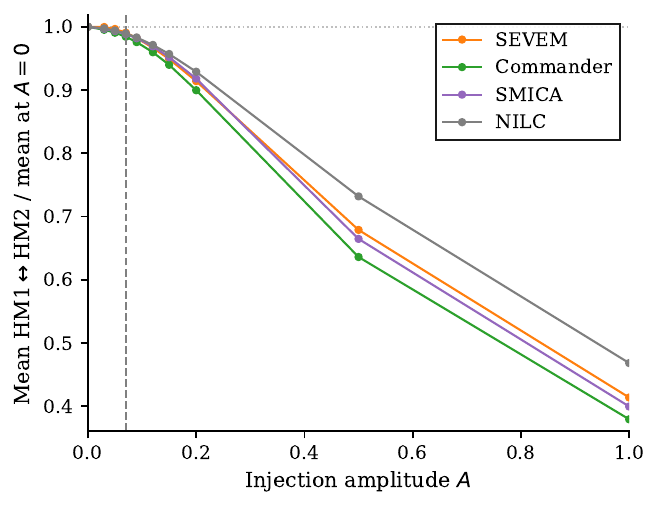}
\caption{Proportional response to injected dipolar modulation for the four Planck component-separation methods. Each curve shows the mean HM1$\leftrightarrow$HM2 angular distance divided by that method's isotropic baseline ($A = 0$). Commander (green) shows the largest fractional reduction. NILC (grey) shows the weakest response. SEVEM (orange) and SMICA (purple) fall between. The vertical dashed line marks $A = 0.07$.}
\label{fig:injection_normalised}
\end{figure}

To isolate the proportional response, we divide each method's mean angular distance at amplitude $A$ by its own isotropic baseline (the mean at $A = 0$). This division-normalized curve (figure~\ref{fig:injection_normalised}) removes the baseline differences and reveals the fractional tightening of half-mission agreement due to the injected signal. Commander shows the largest proportional response across the amplitude range: at each amplitude, its half-mission angular distance decreases by the largest fraction relative to its own baseline. NILC shows the weakest response, consistent with its known excess large-scale polarization noise. SEVEM and SMICA fall between these extremes.

We adopt Commander as the fiducial pipeline on this basis. At $A = 0.07$, the directional consistency check is noise-limited across all four methods: the fractional reduction in angular distance is small relative to the isotropic baseline scatter. The test becomes informative at amplitudes $A \gtrsim 0.5$ for Commander and at progressively higher amplitudes for the other methods (figure~\ref{fig:injection_normalised}).

\section{Results}
\label{sec:results}

\subsection{E-mode polarization}
\label{sec:ee_results}

The deconvolution ablation (section~\ref{sec:ablation}) demonstrates that E--B leakage correction changes the recovered signal. Applying the LVE without deconvolution to the Commander E-mode half-mission maps yields dipole directions more than $90^\circ$ from the temperature HPA direction. The angular distance between them is consistent with isotropic simulations. Without leakage correction, the recovered directions are dominated by mask geometry and galactic contamination residuals. This is the expected outcome and confirms that some form of E--B correction --- whether coupling matrix deconvolution or inpainting, as employed by \citet{GimenoAmo2023} --- is required for polarization HPA analysis on incomplete skies (table~\ref{tab:lbe_lve_comparison}).

\begin{table}
\centering
\caption{Effect of E--B leakage correction: LBE (with deconvolution) versus LVE (without deconvolution) for Commander temperature and E-mode polarization.}\label{tab:lbe_lve_comparison}
\footnotesize
\resizebox{\columnwidth}{!}{%
\begin{tabular}{lccccc}
\hline
Band & Estimator & HM1 direction & HM2 direction & Ang.\ dist. & p-value \\
\hline
TT & LBE & $(223.2^\circ, -36.6^\circ)$ & $(223.0^\circ, -36.6^\circ)$ & $0.1^\circ$ & $6.7\%$ \\
TT & LVE & $(223.1^\circ, -39.5^\circ)$ & $(222.9^\circ, -39.6^\circ)$ & $0.2^\circ$ & $12.3\%$ \\
EE & LBE & $(214.0^\circ, -31.8^\circ)$ & $(208.3^\circ, -24.6^\circ)$ & $8.7^\circ$ & $< 0.33\%$ \\
EE & LVE & $(146.2^\circ, -18.9^\circ)$ & $(269.2^\circ, -38.5^\circ)$ & $101.6^\circ$ & $75.0\%$ \\
\hline
\end{tabular}
}
\end{table}

With coupling matrix deconvolution, the Commander LBE recovers a dipole direction of $(l, b) = (214.0^\circ, -31.8^\circ)$ from HM1 and $(208.3^\circ, -24.6^\circ)$ from HM2. The angular distance of $8.7^\circ$ between them is smaller than in any of the 300 isotropic FFP10 simulations ($p < 0.33\%$; $95\%$ upper bound $p < 1.0\%$). The effect of deconvolution is substantial: the angular distance drops from $101.6^\circ$ ($p = 75.0\%$) to $8.7^\circ$ ($p < 0.33\%$; table~\ref{tab:ee_results}), and the recovered directions shift from more than $90^\circ$ away to the vicinity of the temperature HPA direction.

\begin{table}
\centering
\caption{E-mode polarization (EE) primary results from the Commander LBE at $\ell = 15$--$40$.}\label{tab:ee_results}
\footnotesize
\begin{tabular}{lcccc}
\hline
 & Direction $(l, b)$ & Amplitude & Ang.\ dist. & p-value \\
\hline
HM1 & $(214.0^\circ, -31.8^\circ)$ & $8.6 \times 10^{-5}$ & --- & --- \\
HM2 & $(208.3^\circ, -24.6^\circ)$ & $2.4 \times 10^{-4}$ & --- & --- \\
HM1--HM2 & --- & --- & $8.7^\circ$ & $< 0.33\%$ \\
\hline
\end{tabular}
\end{table}

The same coupling matrix deconvolution yields BB and EB estimates for each patch. For a dipolar modulation of the scalar primordial power spectrum, the expected signal is concentrated in EE~\cite{Namjoo2015}; BB is zero at leading order for scalar modes and affected only at second order with amplitudes equivalent to $r \simeq 0.005$~\cite{Khodagholizadeh2023}; EB vanishes by parity conservation~\cite{Grain2012}. Table~\ref{tab:spectral_diagnostics} shows the spectral diagnostics on Commander. Neither BB nor EB shows directional coherence between half-missions, consistent with this expectation. The absence of directional coherence in BB also serves as a qualitative check on the deconvolution: residual E-to-B leakage from incomplete coupling matrix inversion would elevate BB bandpowers preferentially in patches with complex mask geometry, which could produce spurious structure in the BB bandpower map.

\begin{table}
\centering
\caption{Spectral diagnostics: half-mission directional consistency for Commander EE, BB, and EB bandpowers from the same coupling matrix deconvolution.}\label{tab:spectral_diagnostics}
\footnotesize
\resizebox{\columnwidth}{!}{%
\begin{tabular}{lcccc}
\hline
Spectrum & HM1 $(l, b)$ & HM2 $(l, b)$ & Ang.\ dist. & p-value \\
\hline
EE & $(214.0^\circ, -31.8^\circ)$ & $(208.3^\circ, -24.6^\circ)$ & $8.7^\circ$ & $< 0.33\%$ \\
BB & $(318.5^\circ, -49.4^\circ)$ & $(224.6^\circ, 43.2^\circ)$ & $123.5^\circ$ & $83.3\%$ \\
EB & $(173.0^\circ, 48.9^\circ)$ & $(6.8^\circ, 5.5^\circ)$ & $124.2^\circ$ & $76.3\%$ \\
\hline
\end{tabular}
}
\end{table}

\begin{table*}
\centering
\caption{E-mode directional consistency check across component separation methods. Each method is calibrated against its own matched FFP10 simulation suite (300 simulations each).}\label{tab:compsep}
\footnotesize
\begin{tabular}{lccccc}
\hline
Method & HM1 $(l, b)$ & HM2 $(l, b)$ & Ang.\ dist. & $N_{\rm sim} \leq \mathrm{data}$ & $p$-value \\
\hline
Commander & $(214.0^\circ, -31.8^\circ)$ & $(208.3^\circ, -24.6^\circ)$ & $8.7^\circ$ & $0/300$ & $< 0.33\%$ \\
SEVEM & $(234.5^\circ, -19.3^\circ)$ & $(219.2^\circ, -13.7^\circ)$ & $15.8^\circ$ & $9/300$ & $3.0\%$ \\
SMICA & $(168.3^\circ, -42.8^\circ)$ & $(191.4^\circ, -22.9^\circ)$ & $27.6^\circ$ & $17/300$ & $5.7\%$ \\
NILC & $(271.9^\circ, -60.2^\circ)$ & $(187.2^\circ, -40.7^\circ)$ & $53.1^\circ$ & $76/300$ & $25.3\%$ \\
\hline
\end{tabular}
\end{table*}

\begin{table*}
\centering
\caption{Cross-spectrum (HM1$\times$HM2) directional consistency check. For each method, the cross-spectrum dipole direction is compared to the independent HM1 and HM2 auto-spectrum directions. The simulation mean reports the typical cross$\leftrightarrow$auto angular separation across 300 isotropic FFP10 realisations. $N \leq$ data: number of isotropic simulations with cross$\leftrightarrow$auto separation at most as small as the data value. The cross-spectrum estimate is free of noise bias by construction; the quoted significances remain calibrated against FFP10 simulations (section~\ref{sec:cross_spectrum}).}\label{tab:cross_spectrum}
\begin{tabular}{lccccccc}
\hline
Method & Cross $(l, b)$ & Cross$\leftrightarrow$HM1 & $N \leq$ data & Cross$\leftrightarrow$HM2 & $N \leq$ data & Sim mean HM1 & Sim mean HM2 \\
\hline
Commander & $(214.4^\circ, -43.2^\circ)$ & $11.4^\circ$ & $6/300$ & $19.2^\circ$ & $36/300$ & $68.1^\circ$ & $67.9^\circ$ \\
SEVEM & $(216.0^\circ, -15.0^\circ)$ & $18.2^\circ$ & $27/300$ & $3.4^\circ$ & $1/300$ & $67.8^\circ$ & $64.7^\circ$ \\
SMICA & $(186.6^\circ, -36.2^\circ)$ & $15.5^\circ$ & $13/300$ & $14.0^\circ$ & $12/300$ & $68.1^\circ$ & $71.1^\circ$ \\
NILC & $(258.7^\circ, -64.5^\circ)$ & $7.5^\circ$ & $5/300$ & $46.2^\circ$ & $99/300$ & $71.2^\circ$ & $72.2^\circ$ \\
\hline
\end{tabular}
\end{table*}

Table~\ref{tab:compsep} reports the directional consistency check across all four component-separation methods. The four methods form a gradient from Commander ($p < 0.33\%$) through SEVEM ($3.0\%$) and SMICA ($5.7\%$) to NILC ($25.3\%$). This ordering broadly tracks the injection responsiveness ranking established in section~\ref{sec:injection_comparison}: Commander is the most responsive and produces the tightest result; NILC is the least responsive and shows no significant agreement. SEVEM and SMICA are close in both rankings, though their relative ordering differs. The dispersion is expected: component separation pipelines differ in their treatment of large-scale polarized foregrounds and instrumental systematics, producing different residual contamination at low multipoles. The overall gradient --- Commander and SEVEM tightest, NILC weakest --- is broadly consistent with that reported by \citet{Planck2018VII} and \citet{GimenoAmo2023}.

\begin{figure*}
\centering
\includegraphics[width=0.9\textwidth]{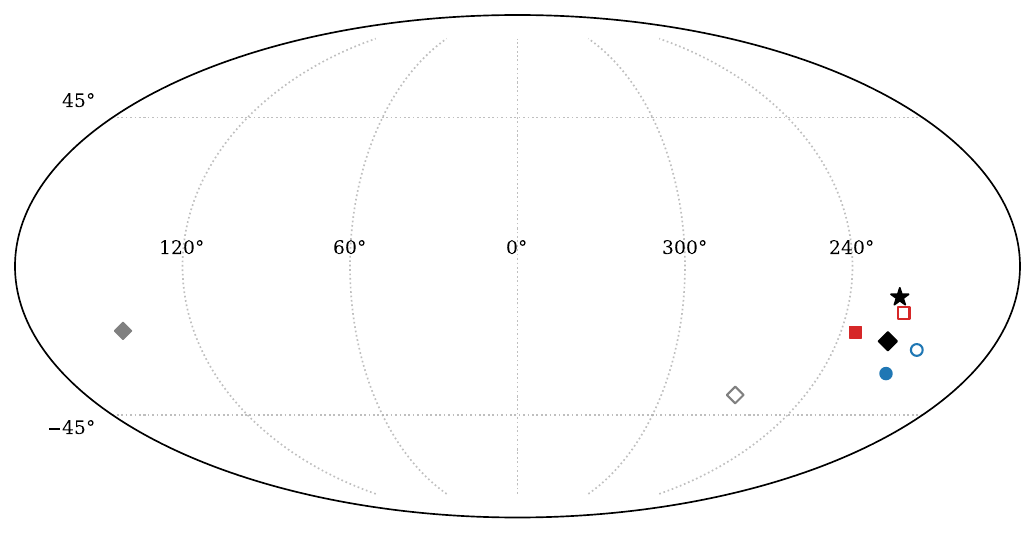}
\caption{Recovered dipole directions in galactic coordinates (Mollweide projection). Blue filled circle: Commander LBE HM1. Blue open circle: Commander LBE HM2. Grey filled diamond: Commander LVE HM1. Grey open diamond: Commander LVE HM2. Red filled square: SEVEM LBE HM1. Red open square: SEVEM LBE HM2. Black star: \citet{GimenoAmo2023} Commander PR3 E-mode direction $(222^\circ, -9^\circ)$. Black diamond: temperature HPA direction $(221^\circ, -22^\circ)$ (\citealt{Planck2018VII}). The LBE directions group near the temperature HPA direction, while the LVE directions (without deconvolution) scatter away from the HPA axis.}
\label{fig:direction_sky}
\end{figure*}

The Commander HM1 direction of $(214.0^\circ, -31.8^\circ)$ lies $11.6^\circ$ from the temperature HPA direction $(221^\circ, -22^\circ)$. The comparison with the inpainting-based results of \citet{GimenoAmo2023} is discussed in section~\ref{sec:comparison} (figure~\ref{fig:direction_sky}).

The dipole amplitude from individual Commander half-missions is consistent with the isotropic simulations: HM1 has an amplitude p-value of $83.3\%$ and HM2 of $11.7\%$. The directional consistency check is sensitive to the coherent signal even when the amplitude is individually unremarkable (section~\ref{sec:why_direction}).

\subsection{Cross-spectrum check}
\label{sec:cross_spectrum}

The auto-spectrum analysis of the preceding section estimates bandpower from each half-mission map individually (HM1$\times$HM1 and HM2$\times$HM2). These auto-spectra contain a noise bias term that is removed by subtracting the FFP10 simulation mean (section~\ref{sec:normalization}). If the FFP10 noise model is locally inaccurate --- producing per-patch noise baselines that differ from those of the real data --- this subtraction could leave residuals that create or suppress a directional signal. Rather than characterise this mismatch directly, the cross-spectrum analysis below anchors the comparison to a direction that is free of noise bias by construction, although the significance of the comparison remains calibrated against FFP10 simulations.

In the HM1$\times$HM2 cross-spectrum, the independent noise terms cancel by construction: the expectation value contains no noise bias term. We compute the cross-spectrum bandpower within each of the 600 patches by passing the HM1 and HM2 $Q/U$ fields as independent inputs to the same \textsc{NaMaster} deconvolution used in the auto-spectrum analysis. A simulation-mean subtraction is applied to remove the isotropic CMB and mask-geometry baseline; unlike the auto-spectrum case, this simulation mean contains no significant noise component, because the independent noise terms have already cancelled before averaging. The resulting cross-spectrum bandpower map is fit with a dipole, yielding a single cross-spectrum direction per component-separation method. The cross-spectrum estimate is therefore free of noise bias. This statement concerns the estimate, not its significance: the cross$\leftrightarrow$auto separations are calibrated against FFP10 simulations, so the quoted p-values retain a dependence on the FFP10 noise model through the null distribution. The auto-spectrum direction used as the reference point likewise retains the noise-subtraction dependence of the auto-spectrum analysis. The cross-spectrum check thus anchors one side of the comparison to a noise-bias-free estimate; it does not remove noise-model dependence from the comparison as a whole.

For Commander, the cross-spectrum direction is $(l, b) = (214.4^\circ, -43.2^\circ)$, which lies $11.4^\circ$ from the HM1 auto-spectrum direction and $19.2^\circ$ from the HM2 auto-spectrum direction. For SEVEM, the cross-spectrum direction $(216.0^\circ, -15.0^\circ)$ lies $18.2^\circ$ from HM1 and $3.4^\circ$ from HM2. For SMICA, the separations are $15.5^\circ$ and $14.0^\circ$. The mean cross$\leftrightarrow$HM1 separation under isotropy is ${\sim}\,68^\circ$ across all methods.

These separations are calibrated against 300 isotropic FFP10 simulations. For all four methods both cross$\leftrightarrow$auto separations are smaller than the isotropic mean of ${\sim}\,68^\circ$; for the three responsive methods (Commander, SEVEM, SMICA) they are reached by only a small fraction of simulations ($0.3$--$12$ per cent), while NILC is consistent with isotropy. The per-method counts and significances are listed in table~\ref{tab:cross_spectrum}. The tightest single agreement is SEVEM's cross$\leftrightarrow$HM2 separation of $3.4^\circ$, which one isotropic simulation in 300 reaches.

The Commander cross$\leftrightarrow$auto separations ($11.4^\circ$ and $19.2^\circ$) are larger than the Commander auto-spectrum HM1$\leftrightarrow$HM2 distance ($8.7^\circ$), suggesting that the extreme tightness of the auto-spectrum result may be partly enhanced by the noise subtraction. For SEVEM, the cross-spectrum direction lies within $3.4^\circ$ of the HM2 auto direction, closely reproducing the auto result. The cross-spectrum check therefore indicates a CMB-level signal across methods while indicating that the specific degree of auto-spectrum tightness --- particularly for Commander --- should be interpreted with the caveat that noise-subtraction effects may contribute.

\subsection{Signal injection characterization}
\label{sec:injection}

We characterise the pipeline's sensitivity through signal injection tests on Commander, in which a known dipolar modulation is applied to FFP10 simulations and the modulated maps are processed through the identical analysis pipeline. For each of the 300 simulations, we inject a modulation at the direction $(l, b) = (45^\circ, 60^\circ)$ --- deliberately offset from the reported HPA direction --- with amplitudes ranging from $A = 0$ to $A = 2$. For each amplitude, we process all 300 modulated simulations and compare the resulting half-mission angular distances to the isotropic null distribution.

\subsubsection{Temperature}

\begin{figure*}
\centering
\includegraphics[width=0.9\textwidth]{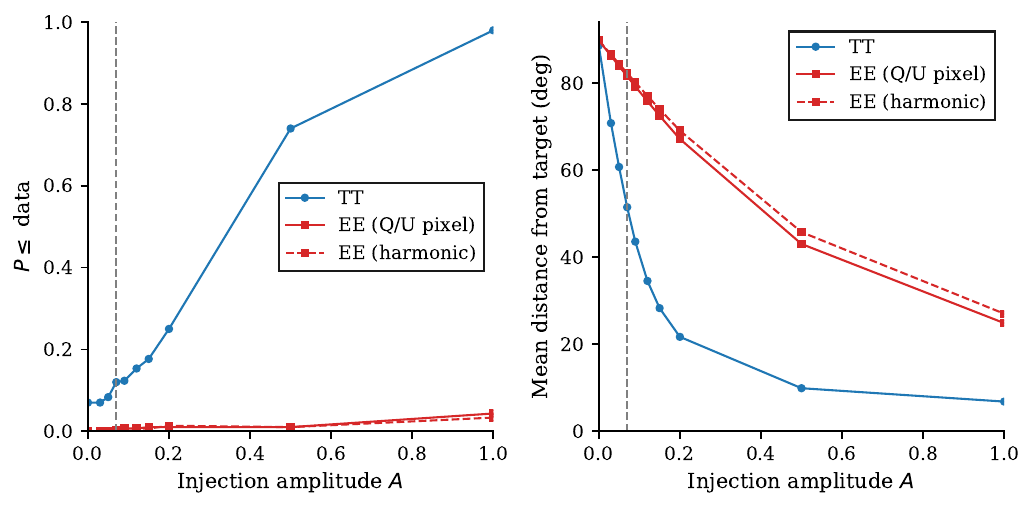}
\caption{Signal injection sensitivity for Commander at amplitudes $A = 0$--$1$. Left panel: $P \leq \mathrm{data}$. Right panel: mean angular distance from the recovered half-mission directions to the injection target (per-simulation mean HM1$\rightarrow$target; same quantities as tables~\ref{tab:injection_tt}--\ref{tab:injection_ee}). TT (blue). EE $Q/U$ injection on Commander Stokes maps (red solid). EE harmonic injection on Commander $E$-modes (red dashed). The vertical dashed line marks $A = 0.07$.}
\label{fig:sensitivity}
\end{figure*}

For temperature, a scalar dipolar modulation $T(\hat{n}) \rightarrow T(\hat{n})[1 + A\,\hat{d} \cdot \hat{n}]$ is applied. Table~\ref{tab:injection_tt} and figure~\ref{fig:sensitivity} summarize the results. At $A = 0.07$, the mean distance from the recovered direction to the injection target is $51.5^\circ$: individual realisations do not reliably recover the injected direction, as the isotropic CMB fluctuations at $\ell = 15$--$40$ produce a fitted dipole of comparable amplitude. Per-realisation direction recovery becomes effective at $A \gtrsim 0.5$, where the mean distance to the target drops to $9.9^\circ$.

\begin{table}
\caption{Commander temperature signal injection results. A scalar dipolar modulation of amplitude $A$ is applied at $(l, b) = (45^\circ, 60^\circ)$ to each of 300 FFP10 simulations. HM$\rightarrow$tgt: mean angular distance from the recovered half-mission directions to the injection target. HM1$\leftrightarrow$HM2: mean half-mission angular distance. $P \leq \mathrm{data}$: fraction of realisations with half-mission angular distance $\leq 0.1^\circ$ (the observed Commander TT value). HM$\rightarrow$tgt and HM1$\leftrightarrow$HM2 are per-simulation means across 300 realisations.}\label{tab:injection_tt}
\centering
\footnotesize
\begin{tabular}{lccc}
\hline
$A$ & HM$\rightarrow$tgt & HM1$\leftrightarrow$HM2 & $P \leq \mathrm{data}$ \\
\hline
$0$    & $88.7^\circ$ & $0.54^\circ$ & $7.0\%$ \\
$0.07$ & $51.5^\circ$ & $0.41^\circ$ & $12.0\%$ \\
$0.09$ & $43.6^\circ$ & $0.35^\circ$ & $12.3\%$ \\
$0.20$ & $21.7^\circ$ & $0.19^\circ$ & $25.0\%$ \\
$0.5$  & $9.9^\circ$  & $0.08^\circ$ & $74.0\%$ \\
$1.0$  & $6.8^\circ$  & $0.04^\circ$ & $98.0\%$ \\
\hline
\end{tabular}
\end{table}

\begin{table*}
\centering
\caption{Commander E-mode polarization signal injection results. Two methods are compared: pixel-space $Q/U$ modulation and harmonic-space E-mode modulation, both applied at $(l, b) = (45^\circ, 60^\circ)$ to 300 FFP10 simulations. HM$\rightarrow$tgt and HM1$\leftrightarrow$HM2 are per-simulation means across 300 realisations. $P \leq \mathrm{data}$: fraction of realisations with half-mission angular distance $\leq 8.7^\circ$ (the observed Commander EE value).}\label{tab:injection_ee}
\begin{tabular}{lcccccc}
\hline
 & \multicolumn{3}{c}{$Q/U$ pixel} & \multicolumn{3}{c}{E-mode harmonic} \\
$A$ & HM$\rightarrow$tgt & HM1$\leftrightarrow$2 & $P \leq$data & HM$\rightarrow$tgt & HM1$\leftrightarrow$2 & $P \leq$data \\
\hline
0 & $89.4^\circ$ & $82.2^\circ$ & $0.0\%$ & $89.4^\circ$ & $82.2^\circ$ & $0.0\%$ \\
0.07 & $81.1^\circ$ & $80.9^\circ$ & $0.3\%$ & $81.9^\circ$ & $81.0^\circ$ & $0.3\%$ \\
0.09 & $78.8^\circ$ & $80.2^\circ$ & $0.7\%$ & $79.7^\circ$ & $80.4^\circ$ & $0.7\%$ \\
0.50 & $42.7^\circ$ & $52.3^\circ$ & $1.0\%$ & $45.4^\circ$ & $54.8^\circ$ & $1.0\%$ \\
1.0 & $24.5^\circ$ & $31.2^\circ$ & $4.3\%$ & $26.7^\circ$ & $33.5^\circ$ & $3.3\%$ \\
2.0 & $13.5^\circ$ & $16.7^\circ$ & $20.3\%$ & $14.8^\circ$ & $18.2^\circ$ & $17.0\%$ \\
\hline
\end{tabular}
\end{table*}

The mean half-mission angular distance decreases monotonically with injection amplitude, and $P \leq \mathrm{data}$ rises correspondingly (table~\ref{tab:injection_tt}). This validates the pipeline's sensitivity to dipolar modulations. The modest shift at the physical amplitude ($A = 0.07$) reflects the high signal-to-noise ratio of the temperature data: under isotropy, the mean half-mission angular distance is already small, leaving little room for a modulation to further tighten the agreement.

\subsubsection{E-mode polarization: Commander detailed characterization}

Table~\ref{tab:injection_ee} and figure~\ref{fig:sensitivity} report the full Commander injection characterization, including both $Q/U$ pixel-space and E-mode harmonic injection pathways. The harmonic injection modulates the E-mode spherical harmonic coefficients through the $\ell$-to-$\ell \pm 1$ coupling induced by a dipole field. The two methods produce nearly identical results across all amplitudes, confirming that the choice of injection pathway does not affect the outcome and that the limiting factor is the per-patch instrumental noise.

At $A = 0.07$, the mean distance from the recovered direction to the injection target is essentially indistinguishable from the unmodulated value. $P \leq \mathrm{data}$ rises only marginally. Per-realisation direction recovery becomes effective at $A \gtrsim 2$, where the mean distance to the target drops below $15^\circ$.

\subsubsection{Interpretation}

The injection tests characterise the sensitivity of the directional consistency check across signal-to-noise regimes. In temperature, the high per-patch signal-to-noise ratio compresses the null distribution to a small mean angular distance; a $7\%$ modulation tightens this only modestly. In E-mode polarization, the low per-patch signal-to-noise ratio produces a broad null distribution (mean $\sim 82^\circ$); a $7\%$ modulation shifts this negligibly. In both cases, the directional consistency check at Planck sensitivity cannot distinguish a $7\%$ modulation from isotropy in individual realisations --- in temperature because the null distribution is already compressed, in polarization because noise dominates. The check becomes informative in an intermediate regime where the signal measurably tightens a broad null distribution; LiteBIRD's improved per-patch polarization sensitivity will move E-mode measurements toward this regime~\cite{LiteBIRD2022}.

The $p < 0.33\%$ for Commander is calibrated against isotropic simulations and quantifies the probability of the observed directional coherence under the null hypothesis. This calibration is independent of the injection model. The observed result is therefore consistent with two interpretations: a noise fluctuation that occurs rarely under isotropy, or a CMB-level signal that the directional consistency check lacks the per-realisation power to confirm at Planck sensitivity. Distinguishing between these requires either a more powerful test statistic or reduced instrumental noise.

Either interpretation rests on the FFP10 calibration, which itself carries a systematic uncertainty: the width of the null distribution is set by the instrumental noise, so a mismatch between the FFP10 noise model and the true noise would shift it --- the noise level sets how strongly the two half-mission directions scatter, and its distribution on the sky sets where noise-driven directions preferentially point. If the simulations overestimate the E-mode noise power on the relevant scales, the simulated separations are systematically broader than the data would produce under isotropy, and the quoted p-values overstate the significance; this applies most directly to the Commander result, in which no simulation reaches the data separation, so part of the tightness could in principle reflect a noise-model mismatch rather than a CMB-level signal. We do not propagate a noise-model uncertainty into the p-values; the quoted values, including the cross-spectrum significances of section~\ref{sec:cross_spectrum}, are conditional on the FFP10 noise model. Two features bound this uncertainty without removing it: the directional-consistency gradient is reproduced across four component-separation pipelines, each calibrated against its own independently processed FFP10 suite; and the cross-spectrum comparison anchors one side to a noise-bias-free estimate.

\section{Robustness tests}
\label{sec:robustness}

\begin{table*}
\centering
\caption{Robustness of the Commander E-mode directional consistency check to pipeline parameters. Each row varies one parameter from the fiducial configuration ($R = 15^\circ$, $\ell = 15$--$40$, no apodisation). All p-values are calibrated against 300 FFP10 simulations processed through the identical pipeline configuration.}\label{tab:robustness}
\footnotesize
\begin{tabular}{llcccc}
\hline
Parameter & Value & HM1 $(l, b)$ & HM2 $(l, b)$ & Ang.\ dist. & $p$-value \\
\hline
Fiducial & --- & $(214.0^\circ, -31.8^\circ)$ & $(208.3^\circ, -24.6^\circ)$ & $8.7^\circ$ & $< 0.33\%$ \\
Apodisation & $1.5^\circ$ C2 & $(212.9^\circ, -32.6^\circ)$ & $(209.2^\circ, -24.7^\circ)$ & $8.5^\circ$ & $< 0.33\%$ \\
Apodisation & $2.0^\circ$ C2 & $(211.9^\circ, -33.9^\circ)$ & $(210.2^\circ, -24.8^\circ)$ & $9.2^\circ$ & $< 0.33\%$ \\
$\ell$-range & $10$--$40$ & $(206.8^\circ, -26.7^\circ)$ & $(203.7^\circ, -18.6^\circ)$ & $8.6^\circ$ & $1.7\%$ \\
$\ell$-range & $15$--$60$ & $(185.8^\circ, -21.0^\circ)$ & $(208.6^\circ, -23.3^\circ)$ & $21.3^\circ$ & $4.7\%$ \\
$\ell$-range & $15$--$80$ & $(192.2^\circ, -20.9^\circ)$ & $(209.5^\circ, -20.8^\circ)$ & $16.2^\circ$ & $1.7\%$ \\
Radius & $12^\circ$ & $(202.9^\circ, -29.2^\circ)$ & $(209.8^\circ, -23.9^\circ)$ & $8.2^\circ$ & $0.3\%$ \\
Radius & $20^\circ$ & $(218.0^\circ, -17.5^\circ)$ & $(202.8^\circ, -26.7^\circ)$ & $16.8^\circ$ & $1.7\%$ \\
\hline
\end{tabular}
\end{table*}

Table~\ref{tab:robustness} summarizes the results for variations of the fiducial Commander pipeline configuration ($R = 15^\circ$, $\ell = 15$--$40$, no apodisation). For each configuration, the p-value is calibrated against FFP10 simulations processed through the identical pipeline.

Applying C2 apodisation at $1.5^\circ$ and $2.0^\circ$ produces negligible changes in both directions and significance (table~\ref{tab:robustness}). The fiducial unapodised configuration was adopted after confirming this insensitivity.

Extending to $\ell = 15$--$60$ and $15$--$80$ yields consistent directions. Varying patch radius from $12^\circ$ to $20^\circ$ produces consistent directions. The fiducial $R = 15^\circ$ balances angular resolution against per-patch signal-to-noise.

Lowering the minimum multipole to $\ell_{\min} = 10$ tests the sensitivity to the lowest modes, which are most affected by cosmic-variance bias~\cite{Ghosh2020}. The result is reported in table~\ref{tab:robustness}.

The significance is calibrated against simulations processed through the identical pipeline for each configuration, ensuring that the p-value is valid independent of the specific parameter choices. The component separation dependence is discussed in section~\ref{sec:ee_results} and table~\ref{tab:compsep}.

\section{Discussion}
\label{sec:discussion}

\subsection{Comparison with previous analyses}
\label{sec:comparison}

The deconvolution-based analysis presented here and the inpainting-based approach of \citet{GimenoAmo2023} address the same challenge --- E--B leakage in local polarization estimation --- through independent strategies. The deconvolution approach corrects leakage through the coupling matrix: for a given patch configuration and input maps, the output is uniquely determined with no stochastic component. The inpainting approach reconstructs the masked field using constrained realisations, yielding results that vary across realisations. The amplitude-based test statistic employed by Gimeno-Amo et al.\ tests the significance of the dipole amplitude directly; the directional consistency check tested here measures half-mission agreement, a different quantity with different statistical power. Both the deconvolution-based directions recovered here and the inpainting-based directions reported by Gimeno-Amo et al.\ on PR3 data point toward the temperature HPA axis, consistent with a common origin. The agreement between two independent leakage correction strategies suggests that the recovered preferred axis is not an artefact of either approach.

\citet{GimenoAmo2025} applied the pseudo-$C_\ell$ estimator in 12 independent sky regions to test whether bandpower dipole directions across multipole bins align with a common axis. No significant alignment was found in E-mode polarization. That analysis tests whether directions align across multipole bins; the directional consistency check tested here measures half-mission agreement at a fixed multipole range. The two analyses use the same underlying estimator; they differ in the spatial sampling (12 independent regions versus 600 overlapping patches) and in the test statistic.

\citet{Ghosh2020} find that $\ell_{\min} \geq 10$ is sufficient for reliable direction recovery in pixel-space polarization analyses; our choice of $\ell_{\min} = 15$ is more conservative, and the robustness test at $\ell_{\min} = 10$ (table~\ref{tab:robustness}) yields consistent directions with a modestly weaker $p$-value (1.7\%). Their $|P|^2$ estimator mixes E and B power inseparably; the pseudo-$C_\ell$ deconvolution separates E from B through the coupling matrix.

\subsection{Cross-method gradient}

The four component-separation methods form a gradient in the directional consistency check: Commander ($p < 0.33\%$), SEVEM ($3.0\%$), SMICA ($5.7\%$), NILC ($25.3\%$). This ordering broadly tracks the injection responsiveness ranking from section~\ref{sec:injection_comparison}: Commander shows the largest proportional response to injected modulations, NILC the weakest. SEVEM and SMICA are close in both rankings, though their relative ordering differs. The overall gradient --- Commander most significant, NILC least --- is consistent with the independently measured sensitivity of each method to CMB-level signals. However, the gradient could also reflect residual foreground power that is preserved by some pipelines and suppressed by others. The directions recovered by Commander and SEVEM --- which employ fundamentally different foreground cleaning strategies --- agree to within $22.6^\circ$, and both point toward the temperature HPA direction rather than toward known polarized foreground structures along the galactic plane. The NILC null result is consistent with this method's known excess large-scale polarization noise rather than with foreground removal of a spurious signal. A definitive separation of cosmological signal from residual foregrounds at these angular scales will require the multi-frequency polarization coverage of LiteBIRD~\cite{LiteBIRD2022}.

\subsection{Multiple comparisons}
\label{sec:look_elsewhere}

The fiducial pipeline (Commander) was selected through the injection comparison of section~\ref{sec:injection_comparison}, which identified Commander as the most responsive method independently of the data p-values. The robustness tests (table~\ref{tab:robustness}) demonstrate stability across pipeline configurations but were not used to select the fiducial. No correction for multiple comparisons is applied, as the additional configurations serve as robustness tests of a single pre-specified analysis rather than independent searches.

\subsection{Directional consistency as a systematics check}
\label{sec:why_direction}

At $\ell = 15$--$40$, the small number of independent modes means that an isotropic CMB realisation produces a dipolar pattern in the bandpower map with amplitude comparable to the expected HPA modulation. Comparing half-mission dipole amplitudes therefore provides little discriminating power at these scales, motivating the use of directional agreement as the test statistic. Both half-missions observe the same CMB realisation; only the instrumental noise differs. If a directional signal is present in the CMB component --- whether from a primordial modulation or from a chance dipolar fluctuation --- it anchors both half-mission directions to the same sky location, producing tighter agreement than noise alone would yield. The check is thus sensitive to the directional coherence of the CMB component rather than its amplitude, accessing information that single-map amplitude tests do not exploit.

An analogy clarifies the logic. Consider detecting a faint astronomical source in two independent exposures. The flux in each exposure is dominated by background noise, rendering a flux-excess test insensitive. However, if a source is present, both exposures will show excess flux at the same position. Positional coincidence between exposures detects the source even when its flux is individually unremarkable. The half-mission directional consistency check exploits a similar principle: a shared CMB signal anchors both half-mission dipole directions to the same sky location. The analogy is imperfect in one important respect: for the faint source, the background is known to be random, so positional coincidence implies a real source. For the CMB, the ``shared component'' includes cosmic variance --- the chance dipolar pattern of the particular CMB realisation we observe. Tight half-mission agreement therefore establishes that the signal resides in the CMB rather than in the noise, but it cannot distinguish a real primordial modulation from a chance CMB fluke. The check is diagnostic, not probative: it rules out independent noise and time-varying systematics as the origin of a candidate signal, but the possibility of a chance CMB realisation remains.

\subsection{Future applications}

The local bandpower diagnostics generalise to any spectrum, multipole range, and data split. It is directly applicable to LiteBIRD~\cite{LiteBIRD2022}, which will provide E-mode maps with substantially lower noise than Planck at the scales relevant to HPA. Detailed forecasts for LiteBIRD sensitivity, including optimization of the pipeline configuration, are left for future work. At LiteBIRD sensitivity, the per-patch signal-to-noise ratio will be high enough for the injection-based calibration to quantify the test's discriminating power across a range of modulation amplitudes.

\section{Conclusions}
\label{sec:conclusions}

We have applied pseudo-$C_\ell$ deconvolution, as implemented in \textsc{NaMaster}~\cite{Alonso2019}, to 600 overlapping patches of the masked sky to construct local bandpower estimates of the CMB E-mode polarization field in Planck PR3 data. The technique is standard; our contribution lies in the diagnostics that the dense patch configuration enables.

An injection-based comparison of the four Planck component-separation pipelines identified Commander as the most responsive to injected dipolar modulations, measured by the fractional reduction in HM1$\leftrightarrow$HM2 angular distance relative to each method's isotropic baseline. We adopted Commander as fiducial on this basis.

Applied to Commander temperature data at $\ell = 15$--$40$, the pipeline recovers the temperature HPA dipole direction with $0.1^\circ$ half-mission agreement, validating the pipeline on a known signal.

In E-mode polarization, removing the deconvolution step causes the recovered directions to shift away from the HPA axis and the half-mission agreement to become consistent with isotropy, confirming that E--B leakage correction is essential. With deconvolution, the Commander directional consistency check gives $p < 0.33\%$ (0/300). SEVEM gives $3.0\%$, SMICA $5.7\%$, and NILC $25.3\%$. The spectral decomposition yields EE directional coherence with null BB and EB, consistent with the expectation for a scalar primordial modulation~\cite{Namjoo2015}. A cross-spectrum (HM1$\times$HM2) check, free of noise bias though still calibrated against FFP10 simulations, corroborates the directional coherence.

We do not interpret this as a detection. The directional consistency check cannot distinguish a real primordial modulation from a chance dipolar realisation in the shared CMB. All quoted p-values are calibrated against FFP10 simulations and are therefore conditional on the accuracy of the FFP10 noise model. Signal injection tests establish that the check is noise-limited at the physical HPA amplitude ($A \approx 0.07$) across all pipelines: a $7\%$ modulation is indistinguishable from the null in individual realisations. The cross-pipeline gradient warrants caution.

The convergence of these diagnostics --- half-mission consistency among the responsive methods, agreement with the temperature HPA direction, cross-spectrum corroboration anchored by a noise-bias-free direction, and EE-only spectral coherence --- is what the framework is designed to evaluate. The observed pattern on Planck data warrants follow-up with LiteBIRD~\cite{LiteBIRD2022}, which will provide the per-patch sensitivity needed to move E-mode measurements into the regime where the directional consistency check has greatest discriminating power.

\section*{Acknowledgements}
We thank the anonymous referee for a detailed and constructive report that substantially improved this paper. We acknowledge the use of the Planck Legacy Archive, the \textsc{healpix}~\citep{Gorski2005} and \texttt{healpy}~\citep{Zonca2019} packages, and the \textsc{NaMaster}~\citep{Alonso2019} code. This work is based on publicly available data from the Planck Legacy Archive (\url{https://pla.esac.esa.int}). The presented results are based on observations obtained with Planck, an ESA science mission with instruments and contributions directly funded by ESA Member States, NASA, and Canada. The analysis code is available at \url{https://github.com/ralynch4916/LBEHPA}. Claude (Anthropic) and Cursor were used to assist with code development and manuscript preparation.
\section*{Data Availability}
The data underlying this article are publicly available from the Planck Legacy Archive at \url{https://pla.esac.esa.int}. The analysis code is available at \url{https://github.com/ralynch4916/LBEHPA}.

\bibliographystyle{mnras}
\bibliography{refs}

\bsp
\label{lastpage}
\end{document}